\input{aipcheck}

\documentclass[
    ,draft            
    ,numberedheadings 
  ]
  {aipproc}

\layoutstyle{6x9}

\begin{document}

\def\aj{AJ}
\def\apj{ApJ}
\def\apjs{ApJS}
\def\apjl{ApJL}
\def\aap{A\&A}  
\def\mnras{MNRAS}
\def\pasp{PASP}
\def\araa{ARAA}
\def\nat{Nature}
\def\pasj{PASJ}

\title{The supernova rate: a critical ingredient and an important tool}


\author{Filippo Mannucci}{
  address={INAF-OAA, Largo E. Fermi 5, 50125 Firenze, Italia}
}

\begin{abstract}
In this review I summarize the role of supernova rate as a 
critical ingredient of modern astrophysics, and as an important tool 
to understand SN explosions.
Many years of active observations and theoretical modeling have 
produced several important results. In particular, linking SN rates
with parent stellar populations has proved to be an important strategy.
Despite these advances, the situation is far from clear, 
in particular for the SNe Ia. 
\end{abstract}

\maketitle


\section{Why bother with how many SNe are exploding}

SN rate has two distinct roles in modern astrophysics. On one side, it is 
a critical ingredient to be used in any model of galaxy formation
and chemical enrichment. 
On the other side, SN rates are also a tool to investigate 
the nature of the exploding stars.

\subsection{An ingredient}

There are many fields were SNe play a key role.
1. Both core-collapse and thermonuclear SNe are the main producers of heavy
elements (e.g., \cite{Matteucci86}).
The chemical enrichment of each galaxy
is determined by the SN rate as a function of galaxy age, while
the cosmic SN rate as a function of redshift is leading the chemical 
evolution of the universe as a whole \cite{Matteucci06,Scannapieco05a}.
2. Core-collapse (CC) SNe are believed to be the main producer of dust at high 
redshifts \cite{Maiolino04a,Maiolino04b,Rho09}.
3. Both types of SNe could contribute and even dominate the feedback processes
needed for galaxy formation and to explain
the ubiquitous presence of outflows in star forming galaxies 
(e.g., \cite{Mori04}).
4. If a reasonable calibration can be obtained, and if dust extinction
can be estimated and controlled (e.g., \cite{Mannucci07b}), than 
the CC rate can also be used to estimate to star formation (SF) density
and its evolution with redshift. Many indicators are now available (see, 
for example, \cite{Hopkins06,Mannucci07a}),
each one with different uncertainties and biases, and it is therefore 
important to compare different, independent result 
\cite{Dahlen04,Dahlen08,Botticella08}.

Summarizing, many fields of astrophysics need to know how many SNe
are exploding at each redshift, in what environments, and what are the 
properties of the ejected material. 
The importance of a good modeling of SN rate should not be underestimated. 
Naive approximations, such as that all SN Ia explode after 1 Gyr, are likely to 
produce completely wrong results.

\subsection{A tool}

In the same time, SN rates are also an important tool to investigate the 
nature of the exploding systems.
While the evolution of SN photometry and spectra
 and the stratification of the chemical
elements can constrain the explosion mechanism, 
SN rates are crucial to constraint the progenitors.
For example, soon after the introduction
of the distinction between ``type I'' and ``type II'' SNe \citep{Minkowski41},
\citet{van-den-Bergh59} used the frequency of type I and type II SNe
to investigate the progenitors.

CC SNe are considered to be due to the gravitational collapse of
very massive stars, M$>8$M$_\odot$,
although how massive is still to be defined.
In principle, this uncertainty in the mass range can be solved or reduced
by measuring good rates and the initial mass function (IMF) of the parent
population \cite{Botticella08,Greggio08}.

The situation for SNe Ia is more complex. These SNe are
considered to be due to the thermonuclear explosion 
of a C/O white dwarf (WD).
Such a conclusion follows from a few fundamental
arguments: the explosion requires a degenerate star, such as a white dwarf;
the presence of SNe Ia in old stellar systems means that at least some of
their progenitors must come from old, low-mass stars; the lack of hydrogen
in the SN spectra requires that the progenitor has lost its outer envelope;
and, the released energy
per unit mass is of the order of the energy output of the 
thermonuclear conversion of carbon or oxygen into iron. 
Considerable uncertainties about the explosion model
remain within this broad framework, such as the structure and
the composition of the exploding WD (He, C/O, or O/Ne), its mass at
explosion (at, below, or above the Chandrasekhar mass) and flame 
propagation (detonation, deflagration, or a combination of the two).

Large uncertainties also remain on the nature of the progenitor system.
Usually, a binary system is considered, with the WD dwarf accreting mass
either from a non-degenerate secondary star (single-degenerate model, SD) 
or from a secondary WD (double-degenerate model, DD).
The evolution of the binary system
through one or more common envelope phases, and
its configuration 
at the moment of the explosion are not known (see \cite{Yungelson05} for a review).
Single-stars models are also possible  \citep{Tout05,Maoz08a},
and current observations are unable to solve the problem \citep{Maoz08b}.

\section{The delay time distribution}

The key quantity to relate type Ia SN rate to the parent stellar
population is the delay time distribution (DTD), i.e., the distribution 
of the delays between the formation of the progenitor system and its 
explosion as a SN. In general, deriving an expected DTD from a progenitor model
is not an easy task because many parameters 
are involved, such as the initial distribution of  orbital parameters in the 
binary system, the distribution of the mass ratio between primary and
secondary star, the efficiency of mass loss during the common envelope phase,
the efficiency of mass transfer from one star to the other, the 
amount of mass retained by the primary star during accretion. 
Also the uncertainties in the
explosion model play a key role: for example, it is not known if it is
necessary to reach the Chandrasekhar mass to start the explosion, or if it is
enough to be in the sub-Chandrasekhar regime.

Starting from the '80s, several authors have computed 
the expected DTD for SD and DD systems
\cite{Greggio83,Tornambe86,Tornambe89,Tutukov94,Ruiz-Lapuente95,Han95,
Yungelson00,Matteucci01,Belczynski05,Greggio05,Yungelson05,
Chen07,Kobayashi08,Hachisu08a,Hachisu08b}. 
Different models often obtain very different results.
In some cases, all the explosions are concentrated in a very
narrow range of delay time (for example, in the SD
Chandrasekhar-mass model by \citet{Yungelson00} all the SNe explode between 
0.6 and 1.5 Gyr).
In other cases, the explosion occurs at any delay time (from 25 Myr to 12 Gyr,
in the  DD model by \citet{Yungelson00}); 
in some models, all happens soon after the
formation (within 1 Gyr for the SD model by \citet{Belczynski05} ), in other cases
the first SNe explode after a very long time (more than 10 Gyr for the 
semidetached double white dwarf model by \citet{Belczynski05} ); 
some distributions are smooth \cite{Greggio05}, some
others have multiple peaks \cite{Belczynski05}.

The observed SN rate is the convolution of the DTD with the past SF
history of the galaxies. This latter function also determines the stellar
population. As a consequence, studying the SN rates in different parent galaxies
can put strong constraints on the DTD.

\section{50 years of optical observations}

The measured rates of SNe both at low and at high redshifts are based on
optical observations because only at these wavelengths the current
instrumentation has sufficient field coverage, 
spatial resolution, and sensitivity
to detect large numbers of SNe within a reasonable observing time.

In the local universe (z$<$0.1), the rates most commonly used
have been computed by \citet{Cappellaro99} 
and \citet{Mannucci05}. Both works are based
on a SN sample defined by
\citet{Cappellaro97} from a 
a compilation of a few visual and photographic searches. 
The ongoing LOSS SN search is expected to produce a new set of 
SN rates in a short time 
(W. Li et al., in preparation), based on a homogeneous set of several hundreds 
of SNe detected in ten years of CCD searches.

Many past and ongoing searches have been designed to discover distant SNe, 
up to z$\sim$1.5, and measure the evolution of the SN rate
\cite{Pain96,Mannucci99,Hardin00,
   Pain02,Madgwick03,Tonry03,Blanc04,Dahlen04,Strolger04,
   Cappellaro05, Barris06,Neill06,Strolger06,Sullivan06,Poznanski07b,
   Botticella08,Kuznetsova08,Dahlen08,Horesh08,Graham08,Dilday08}.
Despite this large effort, significant uncertainties remain.
On one hand, the rates observed by some groups are not consistent
with other results within
the estimated errors, meaning that at least part of them are affected
by systematic errors that are not well understood. 
On the other hand, rates are
derived from the observed number of SNe after a long list of assumptions
(luminosity, light curve, dust extinction, sensitivity, 
spatial distribution of SN within their parent galaxy, colors, and so on).
Usually, it is assumed that these parameters have no evolution with 
redshift, and this can introduce large uncertainties.
For example, several searches for Ia SNe at high redshift assume that
the moderate average extinction observed locally remains the same at 
any redshift.
If, as it is expected, average extinction actually increases with distance, 
this is likely to introduce an underestimate of the rates at high redshifts. 
Also, the above-mentioned assumptions are calibrated on the local sample 
of SNe, dominated by relatively quiescent galaxies. 
At high redshifts, dusty starburst like LIRG and ULIRG 
become the dominant contribution to star formation,
and SNe might have different properties
\cite{Mannucci07b}.

The rates observed at high redshifts, produced by the convolution of 
the DTD with the cosmic SF history, can in principle be used to constrain 
the DTD.
Actually, large uncertainties are present both in the observed rates and in the
cosmic star formation history, and the combination of these two 
uncertainties makes it impossible to derive meaningful DTDs
\cite{Forster06b,Oda08,Blanc08}. 
The redshift evolution of the rates can still be 
used to put additional constraints, but only when other observations are
considered \cite{Mannucci06}, as explained below.

\smallskip

Galaxy clusters play a special role for the study of galaxy evolution and SNe.
These 
regions are particularly overdense, and most of the galaxies are early-type.
Their stellar populations are usually
considered to be simpler than in the field and dominated by old stars. 
Constraints on the DTD
of Ia SNe at late times can be obtained by measuring cluster SN rates.  
Clusters are also very important to derive information on the chemical 
evolution of the universe, because they
retain all the metals lost by the individual galaxies.
For this reason, the measured SN rate can be compared with the 
total amount of SNe ever exploded in the cluster 
as measured by the integrated
metallicity \cite{Scannapieco05a,Matteucci06,Calura07,de-Plaa07}. 
Finally, the large number of galaxies present in a relatively 
small volume can increase the efficiency of SN detection.
For all these reasons,
a significant effort was put in looking for SNe in galaxy clusters
\cite{Crane77,Barbon78,Caldwell81,Norgaard89,Gal-Yam02,Gal-Yam03,
    Maoz04,Germany04,Sharon07,Sand08,Mannucci08a,Gal-Yam08}. 
The number of detected SNe is still low, and more observational work is 
needed in this field.

\section{Observations at longer wavelengths}

In the last few years several attempt were made to avoid the limitations
of the optical observations and detect SNe in dusty environments.
Several searches targeted the near-IR range, up to 2.2$\mu$m 
\cite{van-Buren89,Grossan99,Bregman00,Mattila01,Morel02,Maiolino02,
       Mannucci03,Alonso03,Mattila04,Mattila07,Cresci07,Kankare08}
and obtained a few detections. The rates measured by \citet{Mannucci03}
show that IR observations 
can detect many more SNe in starburst than optical observations, but
also that most of the expected SNe are still missing. 
This is probably due to the presence of high dust column densities,
preventing the detection of many SNe, or by the dominance on nuclear
starburst.
Ongoing and future projects, based on the ESO instruments HAWK-I and VISTA ,
are expected to produce a larger sample of IR-detected SNe, and study 
their properties.

Significant results, albeit for a small number of galaxies, 
have been obtain at radio wavelengths 
\cite{Smith98,Lonsdale03,Rovilos05,Lonsdale06,Parra07}.
At these frequencies, interferometry allows for very high spatial resolutions, 
and even the inner parts of the starburst are transparent. 
As a result, SNe can be detected in deep images even if 
they explode in the nucleus of the parent galaxy
and are completely enshrouded in dust.
It is difficult to use these results to obtain rates, as the properties 
of SNe in such dense environments are not well known. Nevertheless, these
observation show that a significant fraction of SNe are missing from 
any existing searches. These SNe could be a small fraction 
($\sim$10\%) in the local universe, but are expected to dominate 
the rates at z$>$1 \cite{Mannucci07b}.

The IR space-based telescope Spitzer is actively used to study young SNe 
and SN remnants. Probably it will play an important role also
in detecting new SNe, as the large difference in sensitivity with previous 
instruments is expected to compensate the limited spatial resolution.

\section{Open problems with SNe Ia}

\begin{figure}
  \includegraphics[width=7cm]{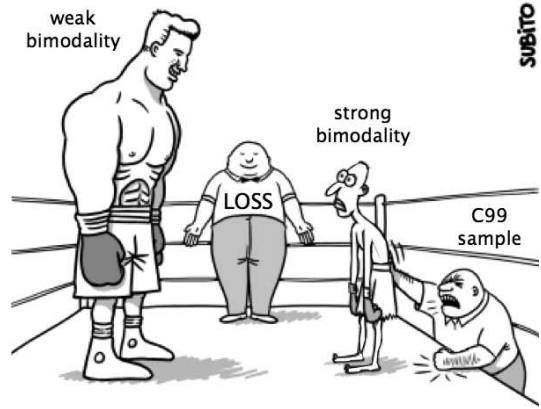}
  \caption{{\em Strong} and {\em weak} bimodality are confronting each other.
The \citet{Cappellaro99} sample supports the strong bimodality, 
and the LOSS sample is the referee. Actually, strong bimodality includes 
the weak one.}
\end{figure}
%

\subsection{The {\em weak} bimodality in type Ia SNe}

Most computations of the DTD for Ia SNe have shown
that binary star 
models naturally predict that these systems
explode from progenitors of very different ages,
from a few 10$^7$ to 10$^{10}$ years. The strongest observational
evidence that this is the case
was provided by \citet{Mannucci05} who analyzed the SN rate per unit stellar
mass in galaxies of all types. We found that the
bluest galaxies, hosting the highest star formation rates (SFRs), 
have SN Ia rates about 30 times larger than those in the reddest, 
quiescent galaxies. 
The higher rates in actively star-forming galaxies imply that
a significant fraction of SNe must be due to young stars,
while SNe from old stellar populations are also
needed to reproduce the SN rate in quiescent galaxies. This lead
\citet{Mannucci05} to introduce the simplified 
two component model for the SN Ia rate 
(a part proportional to the stellar mass and another part to the SFR).
These results were later confirmed by \citet{Sullivan06}, while
\citet{Scannapieco05a}, \citet{Matteucci06} and \citet{Calura07}
successfully applied this model
to explain the chemical evolution of galaxies and galaxy clusters.
A more accurate description is based on the Delay Time Distribution (DTD),
which is found to span a wide range of delay time between a few $10^7$ 
to a few $10^{10}$ years \cite{Mannucci06}.
At least 10\% of the SNe must explode on short timescales ($\sim10^8$ yr)
to follow the SFR, and the rest must follow on much longer timescales. 
This is the so-called ``weak'' bimodality.
Recently, \citet{Pritchet08} have shown that SN rate 
and white dwarf formation rate have the same dependence on the SFR, 
confirming a close link between the two effects and the wide distribution of 
delay times.

\smallskip

Such a wide distribution of DTD is consistent with recent results
based on the Subaru SN search by \citet{Totani08}. 
Usually, DTDs are invetigated by convolving them with theSF  history
to reproduce the observed rates.
Totani et al. invert the process and obtain the DTD by deconvolving 
the SF history from the rates. 
The resulting DTD shows a wide power-law
distribution from $10^8$ to $10^{10}$ years, fully consistent with the results
by \citet{Mannucci05}.
The DTD in \citet{Totani08} depends critically on a severe approximation used,
i.e., that the rate observed in a galaxy is related to the DTD  at
the mass-weighted mean stellar age of that galaxy.
This is a risky hypothesis that should be tested: the system is intrinsically 
non-linear, i.e., galaxies with similar mean age but with different 
age distributions can have SN rates that differs by orders of magnitude.
For example, as little as 0.3\% of young (10$^8$ yr) stars added to 
an old ($10^{10}$ yr) galaxy can easily boost the rate by a factor of two
(assuming the ``classical'' DTD by \citet{Matteucci01}).
The galaxy remains old-looking, the mass weighted mean age 
does not change much, and in any case 
the observed rate is not due to the DTD at that age.
More sophisticated procedures of deconvolution of galaxy spectra 
\cite{Rogers07,Tojeiro07,Fritz07} are needed to check this result.

\subsection{The {\em strong} bimodality in type Ia SNe}

\citet{Della-Valle05} studied the dependence of the SN Ia rate 
in early-type galaxies on the radio power of the host galaxies, and
concluded that the higher rate observed in radio-loud galaxies
is due to minor episodes of accretion of gas or capture of small
galaxies. Such events result in both fueling the 
central black hole, producing the radio activity,
and in creating a new generation of stars, producing the increase in the
SN rate.
The difference between radio-loud and radio-quiet galaxies can
be reproduced by the model of early-type galaxy where
most of the stars are
formed in a remote past, about $10^{10}$ years ago, while a small minority
of stars are created in a number of subsequent bursts. A galaxy appears
radio-loud when is observed during the burst, radio-faint soon after,
and radio-quiet during the quiescent inter-burst period (see \cite{Mannucci08b}).
The amount of mass produced during the bursts can be constrained
by using the (B--K) color observed in both populations.
The results show that the last burst created no more 
that 0.3\% in mass of new stars, assuming 
negligible extinction, or 0.5\% when
assuming an average extinction of the new component of $A_V=1$. 
This model is consistent with several recent works showing the presence 
of mergers, dust, neutral gas, molecular gas and recent star formation 
in local early-type galaxies
\cite{Colbert01,Welch03,van-Dokkum05,Sarzi06,Morganti06,Rogers07,
   Annibali07,Kaviraj07,Schawinski07,Jeong07,Combes07,Kaviraj08}, 
see \citet{Sarzi08} for a recent review.

As the timescale of the radio activity is known to be less than $10^8$ yr, the
rate in early-type radio-loud galaxies can be used to constrain the DTD on 
short timescales. 
Other recent, independent results by \citet{Aubourg08} have confirmed 
the presence 
of a significant fraction of Ia SNe exploding on short timescales.
These evidences are best reproduced by introducing a
``strong'' bimodality:
a ``prompt'' component, comprising 20-60\% of all the Ia SNe, explodes
within $10^8$ yrs, while the   ``tardy'' SNe explode on 
much longer timescale, up to an Hubble time.
 
From a theoretical point of view, it is not difficult to create a bimodal
DTD, especially if different channels of production are considered
\cite{Kobayashi08,Hachisu08b,Wang09}.
Nevertheless, it should be noted that the empirical basis of this ``strong'' 
bimodality \cite{Della-Valle05,Aubourg08} are not as strong as for the
``weak'' bimodality \cite{Mannucci05,Sullivan06}, 
and additional observations are needed to 
confirm the result.

\subsection{Evolution of the properties of Ia SNe}

It is well known that the average properties of local
Ia SNe depend on the host galaxy
(e.g., \cite{Hamuy96a,Gallagher05,Altavilla04,Della-Valle05}), 
and this is also observed at high redshifts 
\cite{Howell07,Howell09}. 
This dependence could be due to both age and metallicity
of the progenitor system, and could be mediated by the amount of 
$^{56}$Ni produced
\cite{Hamuy00,Timmes03,Gallagher08,Howell09}. 
The relative fractions of Ia SNe
produced by young and old systems are expected to change with redshift
as the universe becomes younger and the SF activity increases. 
As a consequence, the average properties of the observed SNe 
could change with cosmic time, and this could be
of extreme importance for the cosmological studies \cite{Riess06,Sarkar08}. 
The presence and the importance of such an effect can be tested by
comparing nearby and distance SNe, looking for differences and similarities.
Current results show that the differences are not large
\cite{Conley06,Garavini07,Bronder08,Ellis08,Foley08,Sullivan09},
nevertheless systematic differences at 10 percent level could be
present \cite{Foley08}.\\


\smallskip

In conclusion, linking SN rates and stellar populations is a valuable tool to 
obtain significant results in both fields of galaxy evolution and SN progenitor.
Oncoming improvements in the SN sample are expected to provide a much clearer
picture.
Nevertheless, the real limiting factor now is the poor knowledge of the 
properties of the galaxies usually targeted by the SN searches. 
Much more accurate conclusions will be reached when the parent stellar 
population can be accurately identified, but this will probably need a new class
of ``galaxy-driven'' SN searches.


%


\bibliographystyle{/Users/filippo/arcetri/Papers/aa-package/bibtex/aa}

\bibliography{/Users/filippo/arcetri/bibdesk/Bibliography}

\IfFileExists{\jobname.bbl}{}
 {\typeout{}
  \typeout{******************************************}
  \typeout{** Please run "bibtex \jobname" to optain}
  \typeout{** the bibliography and then re-run LaTeX}
  \typeout{** twice to fix the references!}
  \typeout{******************************************}
  \typeout{}
 }

\end{document}